\documentclass[12pt,a4paper,fleqn]{article}
\usepackage{tikz}
\usepackage{caption}
\usepackage{lscape}

\usepackage{fullpage}
\usepackage{amsmath}
\usepackage{amsthm}
\usepackage{amssymb}
\usepackage{amscd}
\usepackage{plain}
\usepackage{graphicx}
\usepackage{t1enc,epsfig}
\usepackage[mathscr]{eucal}
\usepackage{epstopdf}
\usepackage[german,english]{babel}
\usepackage{stmaryrd}

\usepackage{xcolor}

\usepackage{wrapfig}
\usepackage{float}

\newtheorem{theorem}{Theorem}[section]

\newtheorem{corollary}[theorem]{Corollary}
\newtheorem{definition}[theorem]{Definition}
\newtheorem{example}[theorem]{Example}
\newtheorem{lemma}[theorem]{Lemma}
\newtheorem{proposition}[theorem]{Proposition}
\newtheorem{remark}[theorem]{Remark}
\newtheorem{conjecture}[theorem]{Conjecture}
\numberwithin{equation}{section}

\renewenvironment{proof}{{\bf Proof. }}{\hfill$\rule{1ex}{1ex}$\par\vskip 5 truemm}

\begin{document}

\newcommand{\xmark}{{\usefont{U}{eur}{m}{n}\symbol{"63}}}
\newcommand{\xmarkk}{{\usefont{U}{eur}{m}{n}\symbol{"43}}}

\newcommand{\bthm}{\begin{theorem}}
\newcommand{\ethm}{\end{theorem}}
\newcommand{\bd}{\begin{definition}}
\newcommand{\ed}{\end{definition}}
\newcommand{\bs}{\begin{proposition}}
\newcommand{\es}{\end{proposition}}
\newcommand{\bp}{\begin{proof}}
\newcommand{\ep}{\end{proof}}
\newcommand{\be}{\begin{equation}}
\newcommand{\ee}{\end{equation}}
\newcommand{\ul}{\underline}
\newcommand{\br}{\begin{remark}}
\newcommand{\er}{\end{remark}}
\newcommand{\bex}{\begin{example}}
\newcommand{\eex}{\end{example}}
\newcommand{\bc}{\begin{corollary}}
\newcommand{\ec}{\end{corollary}}
\newcommand{\bl}{\begin{lemma}}
\newcommand{\el}{\end{lemma}}
\newcommand{\bfj}{\begin{conjecture}}
\newcommand{\ej}{\end{conjecture}}


\def\diy{\displaystyle}

\def\be{\begin{equation}}
\def\ee{\end{equation}}

\def\beq{\begin{equation}}
\def\eeq{\end{equation}}

\def\beal{\begin{array}{l}} \def\beac{\begin{array}{c}} \def\bear{\begin{array}{r}}
\def\beacl{\begin{array}{cl}} \def\beall{\begin{array}{ll}}
\def\bealllll{\begin{array}{lllll}}
\def\beacr{\begin{array}{cr}}
\def\ena{\end{array}}

\def\bma{\begin{matrix}} \def\ema{\end{matrix}}
\def\bpma{\begin{pmatrix}} \def\epma{\end{pmatrix}}

\def\bcs{\begin{cases}} \def\ecs{\end{cases}}

\def\hexnumber@#1{\ifcase#1 0\or 1\or 2\or 3\or 4\or 5\or 6\or 7\or 8\or 9\or A\or B\or C\or D\or E\or F\fi}

\def\N  {{\mathchar"0\hexnumber@\msbfam4E}} 
\def\es {{\mathchar"0\hexnumber@\msbfam3F}}
\def\Z{\mathbb Z}
\def\R{\mathbb R}
\def\L{\mathbb L}
\def\bA{\mathbf A}

\title{{\bf The Pirogov-Sinai Theory for Infinite Interactions}}

\author{\bf A. Mazel$^1$, I. Stuhl$^2$, Y. Suhov$^{2,3}$}

\date{}
\footnotetext{

\noindent
$^1$ AMC Health, New York, NY, USA;\;\;
$^2$ Math Dept, Penn State University, PA, USA;\;\;
$^3$ DPMMS, University of Cambridge and St John's College, Cambridge, UK.}

\maketitle

\begin{abstract}
The purpose of this note is to consider a number of straightforward generalizations of the Pirogov-Sinai theory which can be covered by minor additions to the canonical texts. These generalizations are well-known among the adepts of the Pirogov-Sinai theory but are lacking formal references.
\end{abstract}

\section*{Results}

The original Pirogov-Sinai (PS) theory (\cite{PS, S, Z}) considers lattice models with a finite spin space and a finite translation-invariant potential of a finite radius. Below we use \cite{Z} as the main reference, including notation and terminology. Section 1.1 of \cite{Z} describes the class of considered models in the following way:

\medskip\noindent
``Let $\Z^{\nu}$ be a $\nu$-dimensional lattice ($\nu \ge 2$), and let $S$ be a finite set (of "spins"). For any $\Lambda \subset \Z^{\nu}$, denote by $S^{\Lambda}$ the set of all configurations on $\Lambda$. Suppose that some family $\{\Phi_A\}$ of finite interactions (i.e. functions on $S^A$) is given, invariant with respect to shifts in  $\Z^{\nu}$ and with a finite interaction radius $r$ (i.e. such that $\Phi_A \equiv 0$ if ${\rm diam} A > r$).''

\medskip
The first observation is that the assumption ``of finite interactions'' can be easily removed from all considerations in \cite{Z} by replacing the notion of a configuration, i.e., an element of $S^{\Lambda}$, with the notion of an admissible configuration, i.e., an element $\phi \in S^{\Lambda}$ with $\Phi_A(\phi_A) < +\infty$ for all $A \subset \Lambda$. More precisely, suppose that the first paragraph from \cite{Z} is replaced with the following framed text.

\bigskip
\hrule
\medskip
Let $\Z^{\nu}$ be a $\nu$-dimensional lattice ($\nu \ge 2$), and let $S$ be a finite set (of "spins"). For any $\Lambda \subset \Z^{\nu}$, denote by $S^{\Lambda}$ the set of all configurations on $\Lambda$. Suppose that some family $\{\Phi_A\}$ of interactions (i.e. functions on $S^A$ with values in $\R\cup\{+\infty\}$) is given, invariant with respect to shifts in  $\Z^{\nu}$ and with a finite interaction radius $r$, i.e. such that $\Phi_A \equiv 0$ if ${\rm diam} A > r$. Here diam (and also dist) corresponds to the norm $||t|| = \sum_{i=1}^{\nu} |t_i|$, $t \in \Z^{\nu}$. Accordingly, given $\Lambda \subset \Z^{\nu}$, we set 
$\partial \Lambda = \{t \in \Lambda:\; {\rm dist}(t, \Lambda^c) = 1 \}$ and $\partial^{(n)} \Lambda = \{t \in \Lambda:\; {\rm dist}(t, \Lambda^c) \le n \}$.  The interactions $\Phi_A$ are allowed to take the value~$+\infty$ implying that some configurations from $S^{\Lambda}$ are excluded from the consideration. The remaining part of $S^{\Lambda}$ consists of {\it admissible} configurations having $\Phi_A(\phi_A) < +\infty$ for all $A \subset \Lambda$. Clearly, in a model without infinite interactions every configuration is admissible. We assume that 
the set of admissible configurations is rich enough in the following sense. For any $\Lambda \subset \Z^{\nu}$ there exists at least one admissible configuration in $\Lambda$. Moreover, any pair of admissible configurations in $\Lambda$ and $\Lambda^c \setminus \partial^{(r)} \Lambda^c$ is the restriction of an admissible configuration in $\Z^{\nu}$ (which is not necessarily unique). From now on all
considered configurations are assumed to be admissible and for brevity we omit the word
admissible. Consequently, $\Phi_A$ refers to the finite part of the original interactions. Finally, when we speak about a  family of Hamiltonians we always assume that all Hamiltonians in the family have a common set of admissible configurations.
\medskip
\hrule

\bigskip
Our claim is that after such a replacement the rest of \cite{Z} remains correct {\it verbatim}. 

\medskip
At this point the reader of the current note may switch to the text of \cite{Z} and continue reading starting from the second paragraph of Section 1.1. To help the reader we provide an intuitive explanation of why the approach adopted in \cite{Z} works for infinite interactions in the same way as it does for finite ones.

The fundamental idea of the PS theory is the reduction of the original model to a so called {\it contour model}.
The reduction is done via a chain of identities, which allows one to represent the partition functions of the original model as the partition functions of the contour model. The advantage of the contour model is that the main interaction between contours is  the requirement that their supports do not overlap. Accordingly, polymer expansion 
techniques can be applied to contour models, providing a detailed information regarding the corresponding probability measures in both finite and infinite volumes.

The reduction to the contour model is done via constructing a one-to-one map between an admissible configuration and a compatible collection of contours. Under this mapping it turns out that the statistical 
weight of the admissible configuration is the product of statistical weights of the corresponding contours. Such ``independence'' of contours is the main prerequisite for applicability of polymer expansions.

The reduction to the contour model is the subject of Sections 1.1-1.6 in \cite{Z}. Under the assumptions specified above, everything there is directly applicable to admissible configurations in the models with finite or infinite interactions. This includes the definition of a $q$-{\it correct point}, the definition of a {\it boundary} of an admissible configuration, the definition of a {\it contour} together with its {\it support} and {\it interior}, and, finally, the definition (1.6) of the contour potential $\Phi(\Gamma)$. Note that, due to the definition used in \cite{Z} (see the first paragraph of Section 1.2), a ground state constant configuration is always admissible. On the contrary, in the case of infinite interactions a constant configurations which is not a ground state can be admissible or not. Also, it is the assumption of richness which implies the existence of non-correct points and, consequently, the lower bound $\exp(c|\Lambda|), c > 0$, for the partition function in a finite $\Lambda \subset \Z^{\nu}$ with an arbitrary boundary condition on $\Lambda^c$. Similarly, the absolute value of the logarithm of the ratio of two partition functions in $\Lambda$ with two arbitrary boundary conditions is bounded from above by $\exp(c|\partial\Lambda|), c > 0$. 

By construction, a contour $\Gamma = ({\rm supp} \Gamma,\, \chi_{{\rm supp} \Gamma})$ is a pair consisting of a connected lattice subset ${\rm supp} \Gamma$ and an admissible configuration $\chi_{{\rm supp} \Gamma}$ in ${\rm supp} \Gamma$. It is important that, again by construction, a configuration that is mapped to a finite compatible collection $\{\Gamma_i\}$ of contours takes a constant ground state value on each connected component of the set $\Z^{\nu} \setminus \left( \cup_i \, {\rm supp} \Gamma_i\right)$. This fact implies the desired independence of contours, and it remains true for models with infinite interactions in the same way as for models with finite interactions. 

After the reduction to the contour model the rest of \cite{Z} (with the exception of Section 3.2) deals solely with contour models. The corresponding considerations are based on the validity of the {\it Peierls condition} (see (1.9), (2.9), (2.28) in \cite{Z}) that is supposed to hold for models with both finite and infinite interactions. Thus, these considerations do not make any distinction between finite and infinite interactions. Note that the Peierls condition in the form (1.41) and (2.31) contains constants $K$ and $L$, respectively, which are independent of the Peierls constant $\tau$ from (1.9), (2.9) and (2.28). Constants $K$ and $L$ emerge from the value $e^{\min(K,L)|{\rm supp} \Gamma|}$ that provides an upper bound for the number of possible contours $\Gamma = ({\rm supp} \Gamma, \chi_{{\rm supp} \Gamma})$ with a given ${\rm supp} \Gamma$. 
The upper bound for the number of all configurations in ${\rm supp} \Gamma$ remains valid for a smaller number of admissible configurations.

Section 3.2 in \cite{Z} deals with an arbitrary admissible boundary condition specified on $\Lambda^c$. The only place in this section which is sensitive to infinite interactions is the lower bound (3.36) which follows from the bound above it. The later is valid because of the richness assumption and boundedness of the finite part of the interactions. Such an estimate is violated for at least some infinite interactions not satisfying the richness assumption.

\medskip
The next generalization comes from the fact that the theory in \cite{Z} is applicable to models with a finite number of ground states where each of them is a {\it constant} configuration. It was assumed without saying that an extension to the case of a finite number of {\it periodic} ground states (as considered in \cite{PS, S}) is straightforward. Here is one way to accomplish such an extension.


Observe that for a finite collection of periodic configurations in $\Z^\nu$ there exists a lattice cube $C$ such 
that each configuration from the collection fits $C$ considered as a torus.
Let $C$ be such a cube for the collection of periodic ground states.
Without loss of generality, assume that the linear size of $C$ is larger than the interaction radius. 
Partition $\Z^{\nu}$ into cubes $C(x)\cong C$, where $x \in l \cdot \Z^{\nu}$, and $l$ is the linear size of $C$. The original spin variable $\phi_y$ is associated with $y \in \Z^{\nu}$. Now, replace the spin space $S$ with the spin space $S^C$ and introduce a spin variable $\chi_x = \phi_{C(x)}$, $x \in l \cdot \Z^{\nu}$. After rescaling by the factor $1 \over l$, lattice $l \cdot \Z^{\nu}$ is transformed into lattice $\Z^{\nu}$, and the interaction radius (for both finite and infinite parts of the interaction) becomes $\nu$, since by construction $\chi_{x'}$ interacts 
with $\chi_{x''}$ iff $C(x')$ and $C(x'')$ belong to the same lattice cube of linear size $2l$. This confines us to the settings of \cite{Z} with $r = \nu$. Note that the admissibility criteria (the infinite part of the interaction) 
forbid some pairs $\chi_{x'}$ and $\chi_{x''}$ when $C(x')$ and $C(x'')$ belong to the same lattice cube of linear 
size $2l$. 
 
A further generalization emerges form the fact that the theory from \cite{Z} is constructed specifically for lattice $\Z^{\nu}$. In fact, it is applicable to a generic lattice $B \cdot \Z^{\nu}$, where $B$ is an invertible 
$\nu \times \nu$ matrix. The norm (and correspondingly the distance) used everywhere in \cite{Z} is defined for 
$t=(t_1,\ldots,t_{\nu}) \in \Z^{\nu}$ as
$||t|| = \sum_{i=1}^{\nu} |t_i|$
so that for $s=(s_1,\ldots,s_{\nu}) \in B \cdot \Z^{\nu}$ we can set
$||s|| = \sum_{i=1}^{\nu} |t_i|$
with $t=B^{-1} s \in \Z^{\nu}$. Consequently, the lattice cubes in $B \cdot \Z^{\nu}$ are understood as the images of the lattice cubes in $\Z^{\nu}$. With such a definition of the norm the subsequent arguments in \cite{Z} (starting with the third paragraph of Section 1.1) remain valid.

Finally, consider a countably infinite point set $\L\subset\R^{\nu}$ which is a union of finitely many disjoint sets congruent to $B\cdot \Z^{\nu}$. We call $\L$ a $\nu$-dimensional {\it periodic point set}. Well-known examples are the two-dimensional honeycomb point set and the three-dimensional hexagonal close-packed point set. The PS theory for lattices can be straightforwardly extended to $\nu$-dimensional point sets with the assumption that the interactions are invariant under the translations by vectors from $B \cdot \Z^{\nu}$. Indeed, consider a parallelepiped $F \subset \R^{\nu}$ spanned by the column vectors of $B$ and partition $\R^{\nu}$ into parallelepipeds congruent to $F$. The set of their centers is congruent to $B\cdot \Z^{\nu}$. Considering an admissible configuration in each element of the partition as a new spin variable reduces the model to the one in $\Z^{\nu}$.

\bigskip\noindent
Thus, we end-up with the the standard PS theory on periodic point sets under the following assumptions:
\begin{description}
\item{-} a finite spin space,
\item{-} a translation-periodic interaction of a finite radius, with values in $\R \cup \{+\infty\}$,
\item{-} a richness of the space of admissible configurations,
\item{-} a finite number of periodic ground states,
\item{-} a Peierls condition for contours.
\end{description}
\medskip\noindent
The specific richness condition used in this note is selected to minimize the scope of corresponding modifications in \cite{Z}. Nevertheless, it is wide enough to cover, for example, models dealing with bounded hard-core geometrical objects in $\R^{\nu}$. As an outcome of the above consideration the following results emerge.

\bigskip\noindent
{\bf Model Assumptions.} 
\begin{description}{\sl 
\item{(i)} For a non-degenerate $\nu\times\nu$ matrix $B$ let $\L\subset\R^{\nu}$, $\nu \ge 2$, be the union of a finite number of disjoint sets each of which is congruent to $B\cdot \Z^{\nu}$. 
Let $F \subset \R^{\nu}$ be the parallelepiped spanned by the column vectors of $B$. Set $\bA = \{A \subset \L:\; A \subset F \}$.

\item{(ii)} Let $S$ be a finite set. For any $\Lambda \subseteq \L$, denote by $\sigma_{\Lambda}$ an element of $S^{\Lambda}$. Let $\{\Phi_A(\sigma_{A})\}$, where $\Phi_A(\sigma_{A}) \in (\R\cup\{+\infty\})$, be a family of functions on $S^A$ defined for all $A \in \bA$. Denote by $\chi_{\Lambda}$ an element of $S^{\Lambda}$ having $\Phi_{A}(\chi_{t+A}) < \infty$ for every  $t + A \subseteq \Lambda$ with $t \in B \cdot \Z^{\nu}$ and $A \in \bA$.

\item{(iii)} For any finite $\Lambda \subset \L$ and any $\chi_{\L}$ the conditional Hamiltonian is defined as
$$H(\chi_{\Lambda}|\chi_{\Lambda^c}) = \sum_{t \in B \cdot \Z^{\nu},\, A \in \bA:\; (t+A) \cap \Lambda \not= \emptyset}\; \Phi_{A}(\chi_{t+A}),$$
where $\chi_\Lambda$ and $\chi_{\Lambda^c}$ are the restrictions of $\chi_{\L}$.

\item{(iv)} The Hamiltonian $H(\chi_{\Lambda}|\chi_{\Lambda^c})$ has a finite number of periodic ground states, and each of them is invariant under shifts by the column vectors of $B$. (Here a ground state is a configuration $\overline \chi_{\L}$ such that $H(\chi_{\Lambda}|\overline\chi_{\Lambda^c}) \ge H(\overline\chi_{\Lambda}|\overline\chi_{\Lambda^c})$ for any finite $\Lambda \subset \L$ and any $\chi_{\L}$ which coincides with $\overline\chi_{\Lambda^c}$ in $\Lambda^c$.)}
\end{description}

\medskip\noindent
{\bf Reduction Lemma.} {\sl Considering $\overline S = S^{\L \cap F}$ instead of $S$ transforms the original model into an equivalent model where $\L = \Z^{\nu}$, $F$ is a unit cube in $\Z^{\nu}$, and all periodic ground states 
are constants.
}

\medskip\noindent
{\bf Simplified Model Assumptions.} 
\begin{description}
{\sl

\item{(i)} 
Let $F$ be the unit cube in $\Z^{\nu}$ 
and set $\bA = \{A \subset \Z^{\nu}:\; A \subseteq F \}$.

\item{(ii)} Let $S$ be a finite set. For any $\Lambda \subseteq \Z^{\nu}$, denote by $\sigma_{\Lambda}$ an element of $S^{\Lambda}$ and let $\{\Phi_A(\sigma_{A})\}$, where $\Phi_A(\sigma_{A}) \in (\R\cup\{+\infty\})$, be a family of functions on $S^A$ defined for all $A \in \bA$. Denote by $\chi_{\Lambda}$ an element of $S^{\Lambda}$ having $\Phi_{A}(\chi_{t+A}) < \infty$ for every  $t + A \subseteq \Lambda$ with $t \in \Z^{\nu}$ and $A \in \bA$.

\item{(iii)} For any finite $\Lambda \subset \L$ and any $\chi_{\Z^{\nu}}$ the conditional Hamiltonian is defined as
$$H(\chi_{\Lambda}|\chi_{\Lambda^c}) = \sum_{t \in B \cdot \Z^{\nu},\, A \in \bA:\; (t+A) \cap \Lambda \not= \emptyset}\; \Phi_{A}(\chi_{t+A}),$$
where $\chi_\Lambda$ and $\chi_{\Lambda^c}$ are the restrictions of $\chi_{\Z^\nu}$.

\item{(iv)} The Hamiltonian $H(\chi_{\Lambda}|\chi_{\Lambda^c})$ has a finite number of periodic ground states, and each of them is a constant function on $\Z^\nu$.}
\end{description}

\medskip\noindent
{\bf Peierls Assumption.} {\sl 

\medskip
Let $\Gamma = ({\rm supp} \Gamma,\, \chi_{{\rm supp} \Gamma})$ be a pair consisting of a finite connected set ${\rm supp} \Gamma \subset \Z^{\nu}$ and $\chi_{{\rm supp} \Gamma}$, where $\chi_{{\rm supp} \Gamma}$ is a restriction of $\chi_{\Z^{\nu}}$, and $\chi_{\Z^{\nu}}$ has the following properties.

\begin{description}
\item{(i)} $\chi_{\Z^{\nu}}$ has a constant ground state value on each connected component of \,$\Z^{\nu} \setminus {\rm supp} \Gamma$. In particular, let ${\rm sign}(\Gamma) \in S$ be the constant ground state value corresponding to the infinite component of \,$\Z^{\nu} \setminus {\rm supp} \Gamma$.

\item{(ii)} For any $t$ belonging to a given connected component of \,$\Z^{\nu} \setminus {\rm supp} \Gamma$
the function $\chi_{U(t)}$ is the restriction of a constant ground state.  Here $U(t)$ is a lattice cube of linear size 
$3$ centered at $t$. 

\item{(iii)} For any $t \in {\rm supp} \Gamma$ and $U(t)$ as above,  function $\chi_{U(t)}$ is not the restriction of a constant ground state.

\item{(iv)} There exists a (Peierls) constant $\tau > 0$ such that for any 
$\Gamma = ({\rm supp} \Gamma,\, \chi_{{\rm supp} \Gamma})$
$$ H(\chi_{{\rm supp} \Gamma}|\chi_{({\rm supp} \Gamma )^c}) - H(\overline\chi_{{\rm supp} \Gamma}|\overline\chi_{({\rm supp} \Gamma)^c})> \tau |{\rm supp} \Gamma|, $$ 
where $|{\rm supp} \Gamma|$ is the number of sites in ${\rm supp} \Gamma$ and $\overline\chi_{\Z^{\nu}} \equiv {\rm sign}(\Gamma)$. 
\end{description}
}

\medskip\noindent
{\bf Richness Assumption.} {\sl 

\medskip
For any $\Lambda \subset \Z^{\nu}$ there exists at least one $\chi_{\Lambda}$. Moreover, any pair $\chi_{\Lambda}$ and $\chi_{\Lambda^c \setminus \partial^{(n)} \Lambda^c}$ is the restriction of some (not necessarily unique) $\chi_{\Z^{\nu}}$, where $\partial^{(n)} \Lambda = \{t \in \Lambda:\; \break\min_{t' \in  \Lambda^c}\, \left( \sum_{i=1}^{\nu} |t_i-t'_i| \right) \le n \}$ and $n$ is a fixed constant independent of $\Lambda$, $\chi_{\Lambda}$ and $\chi_{\Lambda^c \setminus \partial^{(n)} \Lambda^c}$.}

\medskip\noindent
{\bf Theorem.} {\sl Under the Richness, Peierls and Simplified Model Assumptions for $\tau$ large enough  (depending on $\nu$ and $S$) all Lemmas, Theorems and Corollaries of \cite{Z} hold true.}

\medskip\noindent
{\bf Corollary.} {\sl Under the Richness, Peierls and Simplified Model Assumptions for $\tau$ large enough 
(depending on $\nu$ and $S$) the set of constant ground states contains a nonempty subset of so-called 
stable ground states. For each stable ground state with the constant value $\overline\chi$ the following statements are true.
\begin{description}
\item{(i)} The partition function in a finite $\Lambda \subset \Z^{\nu}$ with the boundary condition 
$\overline\chi$
admits a cluster representation (see (2.1) in \cite{Z}) in terms of collections $\{\Gamma_i \}$ with mutually disconnected ${\rm supp} \Gamma_i$ and ${\rm sign}(\Gamma_i) = \overline \chi$.

\item{(ii)} The logarithm of this partition function admits an explicit expression as an absolute convergent sum of statistical weights of polymers that are collections $[\Gamma_j]$ with connected $\cup_j {\rm supp} \Gamma_j$.

\item{(iii)} The thermodynamic limit ($\Lambda \nearrow \Z^{\nu}$) of the correlation function of any finite collection $\{\Gamma_k \}$ admits an explicit expression in terms of polymers $[\Gamma_j]$, implying the 
existence of the  limit Gibbs measure generated by boundary condition $\overline \chi$. The corresponding free energy admits 
an explicit expression in terms of polymers $[\Gamma_j]$.

\item{(iv)} The limit Gibbs measure generated by $\overline \chi$ has an exponential decay of correlations 
and is an extreme one.
\end{description}
\noindent
For each non-stable (metastable) constant ground state used as a boundary condition the corresponding 
limit Gibbs measure  is a mixture of extreme (stable) ones from item (iv). Moreover, any translational periodic limit Gibbs measure is a mixture of extreme ones from item (iv).}


\begin{thebibliography}{10} 

\bibitem{PS}
Pirogov, S.A., Sinai, Ya.G. Phase diagrams of classical lattice systems. {\it Teor. Mat. Fiz.}
{\bf 25} (1975), 1185-1192; {\bf 26} (1976), 61-76.

\bibitem{S} Sinai, Ya.G. {\it Theory of phase transitions: rigorous results.}
Oxford: Pergamon Press, 1982.

\bibitem{Z}
Zahradnik, M. An alternate version of Pirogov-Sinai theory. {\it Comm. Math Phys.} {\bf 93}
(1984), 559-581.

\end{thebibliography}
\end{document}